\documentstyle[jltp,11pt,twoside,amssymb]{article}

\title{Magnetothermal Transport of Oriented Graphite at Low Temperatures}

\author{Konstantin Ulrich and Pablo Esquinazi}

\address{Abteilung Supraleitung und Magnetismus,
Universit\"at Leipzig, Linn\'estrasse 5,\\ D-04103 Leipzig, Germany}

\runninghead{K. Ulrich and P. Esquinazi}{Magnetothermal Transport of
Graphite}
\input psfig
\begin{document}

\begin{abstract}
\vspace*{-9cm}
\begin{center}
To be published in \\
Journal of Low Temperature Physics, Volume 137 3/4, November 2004
\end{center}
\vspace*{5cm} We have studied the magnetic field dependence of the thermal
conductivity $\kappa(T,B)$ of highly oriented pyrolytic graphite samples
at temperatures $0.2$K$ \le T < 10$K and fields 0T$ \le B \le 9$T. The
samples show clear deviations from the Wiedemann-Franz law with a kink
behavior at fields $B \sim 0.1~$T~ near the metal-insulator transition
observed in electrical resistivity measurements.  We further show that the
oscillations in the thermal conductivity at the quantum limit $B > 1~$T,
which are correlated with the Landau quantization observed in Hall
measurements, increase in amplitude with temperature following a $\sim
T^3$ law at $T
> 0.2~$K and show a maximum at $T \sim 6~$K, suggesting that they
are phonons mediated.

\end{abstract}

\maketitle


\section{INTRODUCTION}

One of the authors (P.E.) of this contribution met Siegfried Hunklinger
for the first time in a warm day of August in the year 1983 at the
Institut f\"ur Angewandte Physik II located near the Philosophenweg in the
beautiful city of Heidelberg. Part of the work done in the two and a half
years of postdoc stay was guided by him and Georg Weiss. That research
work done twenty years ago was mainly based on the interaction between
tunneling systems (TS) and conduction electrons in amorphous metals. At
that time an anomalous behavior in the acoustic properties of amorphous
superconductors below the superconducting critical temperature was found,
indicating that the interaction of TS and conduction electrons was not as
simply as the Korringa-based models predicted.\cite{neckel86} After all
these years, the main observations and this subject in general remain
still without clear answers. However, we know  that the applied strain
energy and the interaction {\em between} TS play an important role and
should be taken into account.\cite{esqui04} The work initiated and guided
by Siegfried Hunklinger was stimulating and ambitious. For P.E. it was a
pleasure to work and collaborate with him. We are glad to have here the
possibility to honor his scientific trajectory with a contribution on a
relatively new topic in solid state physics.

An unaware reader may be surprised to see a paper reporting on the
magnetothermal transport of graphite. In fact, graphite is one of
the most studied materials in physics and chemistry and the
literature contains a substantial number of measurements and
theoretical work on the electrical and thermal transport
properties of this semi-metal\cite{kel,dre}. In contrast to the
common believe, however, several transport properties of graphite
are not understood and some theoretical assumptions done in the
past seem now less plausible. The number of open questions
regarding the transport properties of graphite is significant.
From the experimental side this is partially due to the fact that
the quality of the samples we can measure today is much better
than that obtained in the past. From the theoretical side, we know
now that the physics of an electron system in two dimensions
cannot be simply derived from the Fermi liquid theory and an
unusual and  still controversial behavior for the transport
properties is predicted and partially observed experimentally. For
ideal, two-dimensional graphite the situation becomes even more
interesting because of the predictions that at the $K$-points of
the Brillouin zone, the electrons should behave as relativistic,
massless Dirac-Fermions (DF) with a linear dispersion relation
\cite{semenoff,divin,gonzalez}, similar to, e.g., the
quasiparticles (QP) at the gap nodes in the high-temperature
superconductors (HTS).\cite{tsuei}

The subject of this contribution is the thermal conductivity and
its magnetic field dependence of highly oriented pyrolytic
graphite (HOPG) samples. This property and its field dependence
was indeed study in oriented samples of less quality in the past
by Ayache in his thesis \cite{aya} and by Woollam \cite{woo}, but
those experimental curves were apparently never published. The
behavior of the thermal conductivity at the quantum limit (applied
magnetic field normal to the graphene planes $B
> 1~$T) where signs of the quantum Hall effect are observed
\cite{yakovprl}, is of interest as well as its behavior at low fields ($B
\sim 0.1~$T) where a field-driven metal-insulator-like transition (MIT) is
measured in the electrical resistivity.\cite{kempassc,yakovprl} We will
show below that the thermal conductivity is a useful and important
transport property to study the behavior of the QP in graphite. Our main
results reveal that $\kappa(B)$ shows an anomalous behavior at the MIT,
which resembles that expected from the magnetic catalysis (MC) theory
\cite{ferrer,KhveshPRL2001a,GorbarPRB2002} and that the amplitude of the
quantum oscillations in $\kappa(B
> 1~$T) decreases with decreasing temperature following a $T^n$ law with
$n = 3.0 \ldots 3.5$, in clear
contrast to the electrical Hall effect behavior.

\section{PREVIOUS EXPERIMENTAL WORK AND THEORETICAL BACKGROUND}

In a recent paper \cite{ocana03} we have reported on the magnetothermal
conductivity of a HOPG sample at temperatures 5~K $\le T\le$ 20~K for
fields (0T$ \le B \le 9$T) parallel to the $c-$axis. With the measured
longitudinal electrical resistivity we showed that the Wiedemann-Franz law
is not able to explain the large oscillations in $\kappa(B)$ observed in
the high-field regime $(B > 1~$T) where
 a quantum-Hall-like behavior is measured. The temperature dependence of the
oscillation amplitude $\Delta k$ between the minimum (at $B \simeq
3.7~$T) and maximum (at $B \simeq 5.5~$T)  showed a rough maximum
at $\sim 7~$K, i.e. near the limit of the measuring range. Because
the phonon-electron interaction does not appear to provide an
answer for the large oscillations in $\kappa(B)$,\cite{ocana03}
lower temperature measurements are necessary to get more hints on
their origin.

The scientific interest on graphite has been recently renewed by transport
measurements on HOPG
\cite{KopelPSS1999,kempassc,SercheliSSC2002,KempaPRB2002}, which shows a
similar MIT as in two-dimensional (2D) electron (hole) systems (which
takes place either varying carrier concentration or applying a magnetic
field $B$) \cite{AbrahamsRMP2001}. The MIT in graphite is remarkable: the
in-plane electrical resistivity can increase a factor 10 after the
application of a magnetic field of $\sim 0.2~$T perpendicular to the
graphene planes and the temperature dependence changes from metallic-like
($d\rho(T)/dT > 0$) to semiconducting-like ($d\rho(T)/dT < 0$) at a
``critical" field $B_c \sim 0.04 \ldots 0.1~$T.

Theoretical analysis \cite{KhveshPRL2001a,GorbarPRB2002} suggests that the
MIT in graphite is the condensed-matter realization of the magnetic
catalysis (MC) phenomenon \cite{GusyninPRL1994} known in relativistic
theories of (2 + 1) - dimensional DF. According to this theory
\cite{KhveshPRL2001a,GorbarPRB2002}, the magnetic field opens an
insulating gap in the spectrum of DF associated with an electron-hole
(e-h) pairing, below a transition temperature $T_{ce}(B)$. In qualitative
agreement with theory, experiments show a $T_{ce}(B)$ that increases with
field from a field of the order of 50~mT.\cite{yakovadv}

The influence of the MC on the electronic contribution to the thermal
conductivity for a DF system has been calculated in Ref.~13. The MC was
suggested to be the reason for the kink behavior found in the magnetic
field dependence of the thermal conductivity of some high-temperature
superconducting samples.\cite{krishana2,aubin} Experimentally, however,
this kink behavior is not systematically observed\cite{pogo,talden,ando},
probably due to a in-plane anisotropy of the cuprates\cite{vief} added to
the fact that QP-vortices interaction plays a main role in the field
dependence of $\kappa$ in the superconducting state. It seems that HTS
materials are not the best candidates to study the behavior of  DF as a
function of field. The MC theory predicts that above a ``critical" field
$B_c(T)$, $\kappa(B)$ shows a plateau-like behavior being larger than the
conductivity one would obtain from, e.g. the WF law. The reason for this
effective increase of $\kappa(B > B_c)$ in the theory is due to the
decrease of the efficiency of the relevant scattering processes  for the
thermal transport of the excitonic pairs.\cite{ferrer} According to the
authors \cite{ferrer} the appearance of a kink in $\kappa(B)$ is model
independent, being only determined by the critical behavior of the induced
dynamical mass near the phase transition. We may qualitatively expect that
the effect of the MC should be seen when the applied field $B \gtrsim
B_c$, the field needed to generate the dynamical mass at a temperature
$T$, as described in Ref.~13. In a previous experiment\cite{ocana03} we
did not find clear evidence for a kink in $\kappa(T \gtrsim 5$K$,B < 1~$T)
in contrast to what we are reporting here. The results suggest that the
kink is sample dependent and that the intrinsic disorder may play a role.
We remark that for the relativistic spectrum of 2D-QP in graphite the
violation of the WF law is expected simply because the total electrical
current differs from the total heat current for interband excitations that
create quasiholes.\cite{yang}

\section{EXPERIMENTAL DETAILS}

We have studied two HOPG samples: UC-sample from Union Carbide
(rocking curve FWHM $\simeq 0.26^o$), AC-sample from Advanced
Ceramics (rocking curve FWHM $\simeq 0.4^o$). The UC-sample is a
different piece from the same batch of the HOPG sample reported in
Ref.~16. The samples typical length and width were $\sim 1 \ldots
\sim 2~$mm and thickness $\sim 0.1$~mm.

The thermal conductivity measurements at $T \ge 5~$K of the UC-sample were
done in a 4~K cryostat equipped with a 9~T superconducting solenoid. For
the measurement of the temperature gradient (between 100~mK and 200~mK in
$\sim 1~$mm length) we used a previously field- and temperature-calibrated
type E thermocouples with a dc-picovoltmeter \cite{iny}.  The thermocouple
ends were positioned both at the same surface of the sample. A detailed
calibration of the thermocouple as a function of field and temperature was
performed because  the thermopower of our thermocouple is specially
sensitive on the magnetic field below 10~K with a non-simple
dependence\cite{iny,ulrich}. The experimental arrangement was recently
used to study the longitudinal and Hall thermal conductivities of
high-temperature superconducting crystals\cite{oca2}. Our system enables
us to measure $\kappa(B)$ with a relative resolution better than $0.1\%$
above 5~K. The thermal stability was better than 1~mK in the measured
magnetic field and temperature range. The absolute error in the thermal
conductivity was estimated to be $\le 30\%$.

The lower temperature measurements were done in a dilution cryostat (0.2~K
$ \le T \le 4~$K) equipped with a 8~T superconducting solenoid. For these
measurements and due to the necessity of using  small pieces of RuO$_2$
resistances to measure the temperature gradient (the sensitivity of the
tiny thermoelement is not good enough below 5~K) we were forced to use the
AC-sample due to its larger dimensions, although our experience indicates
that larger HOPG samples may show a less clear two-dimensional behavior in
their transport properties due to the internal disorder. The RuO$_2$
thermometer shows a relatively small field dependence and has a good
temperature sensitivity at low temperatures.\cite{ruo2} Four RuO$_2$
thermometers were calibrated before they were attached to the sample and
this calibration was checked during the thermal conductivity measurements.

\section{EXPERIMENTAL RESULTS AND DISCUSSION}
\subsection{Deviations from the WF law at magnetic fields $B \sim 0.1~$T}
\begin{figure}
\vspace{-0.2cm}\centerline{\psfig{file=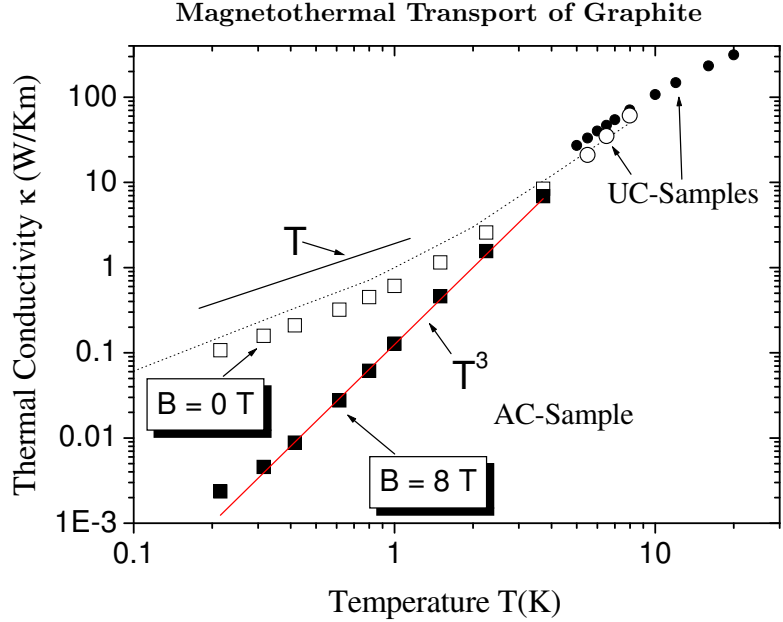,height=3.1in}}
\caption{Thermal conductivity as a function of temperature for two
UC-samples ($\bullet$ from Ref.~16 and $(\circ$) this work) at $B = 0~$T
and for the AC-sample at $B = 0~$T ($\Box$) and at $B = 8~$T
($\blacksquare$). The dashed line represents the measured data for a HOPG
sample taken from Ref.~34.} \label{kt}
\end{figure}

Figure \ref{kt} shows the temperature dependence of the thermal
conductivity of the AC-sample measured in zero field and under 8~T. In the
same figure we show the corresponding data for two UC-samples at $B=0$.
The measured temperature dependence as well as the absolute value agree
with earlier measurements done in HOPG samples (dashed
line).\cite{morelli} We assume that the thermal transport of graphite is
given by two contributions \cite{kel}:
\begin{equation}
\kappa= \kappa_p(T) + \kappa_e(T,B)\,, \label{kappa}
\end{equation}
where $\kappa_p$ is due to the phonons and $\kappa_e$ due to charged
carriers (conduction electrons and/or holes). The field dependence of the
thermal conductivity is given only by the electronic part $\kappa_e(T,B)$,
which can be estimated with the WF relation. Experimental evidence shows
that the field dependence of $\kappa$ decreases with temperature being
negligible at 100~K (($\kappa(B)-\kappa(0))/\kappa(0) < 0.1\%$ at $B =
9~$T), indicating that the phonon contribution does not change with field.
The universal WF-relation relates the electrical resistivity $\rho(T,B)$
with the thermal conductivity due to electrons by
\begin{equation}
\frac{\kappa_e \rho}{T} = L_0 \label{wf}\,,
\end{equation}
 through the universal
constant $L_0 = 2.45 \times 10^{-8}~$W$\Omega$K$^{-2}$.  The relation
(\ref{wf}) holds strictly for elastic or quasielastic electron scattering
and therefore the range of validity is usually set, either at low enough
temperatures where the resistivity is temperature independent (impurity
scattering dominates), or at high enough temperatures where the
electron-phonon scattering is large \cite{am}. For the samples measured in
this work, the temperature dependence of the electrical resistivity
indicates a saturation below 10~K (curves for similar samples can be seen
in Refs.~19,22) and therefore at $T \le 10~$K we expect to be roughly in
the validity range of the WF-law. We stress that within the Fermi liquid
theory the field dependence of $\kappa_e$ should be given by the WF law.
Deviations from the expected field dependence obtained from the measured
electrical resistivity should be taken as a hint for a non-Fermi liquid
behavior, whatever their origin.

Indeed, the estimates of the electronic part from Eq.~(\ref{wf}) using the
measured resistivity are in general in agreement with experimental results
at zero applied field. For example, from Fig.~\ref{kt} we estimate that at
the crossover between the $T^3$- and $T$-dependence of $\kappa$ at $T \sim
1.5$~K and zero field, $\kappa_e \sim \kappa_p \sim 0.5~$W/Km. Taking the
measured resistivity for this sample $\rho(2~$K,0)$\simeq 0.14~\mu\Omega$m
from (\ref{wf}) we obtain $\kappa_e \sim 0.25~$W/Km. Taking into account
the 30\% uncertainty in the absolute value in both $\kappa$ and $\rho$ we
obtain an apparent agreement between the WF law and the experimental
result. On the other hand, the experimental error does not allow us to
rule out the contribution of other QP with a different effective Lorentz
number at $B = 0$.

With magnetic field, however, the apparent agreement clearly breaks down.
As expected from the increase of the electrical resistance with field, the
crossover from $T^3$ to $T$ in the temperature dependence of $\kappa$ is
shifted to lower temperatures, as has been observed experimentally in
earlier work.\cite{chau} From our results, see Fig.~\ref{kt}, this
crossover is observed at $T \sim 0.3~$K at $B = 8~$T and therefore we can
estimate $\kappa_e(0.3$K$,8$T)$ \sim 10^{-3}~$W/Km. The estimate from
Eq.~(\ref{wf}) provides however $\kappa_e \sim 10^{-4}~$W/Km using the
measured resistivity, i.e. much smaller than the estimated value from the
measurements. The $T^3$ dependence found at $B = 8~$T and $T > 0.3~$K
appears to be due to phonon transport limited by grain boundaries, as a
simple estimate indicates. Using the specific heat from literature $C \sim
0.0325 \times 10^{-3} T^3$J/mol K (page 183 in Ref.~3) and an average
sound velocity $v \sim 2 \times 10^4$m/s (page 89 in Ref.~4) we obtain a
phonon mean free path $l \sim 3~\mu$m, which is of the order of the
measured crystallite size for similar samples.\cite{kel,car} Nevertheless
and due to the failure of the WF law we cannot rule out that an enhanced
QP contribution still exists at large fields. This contribution may
influence only weakly the $T^3$ dependence due to phonons.

\begin{figure}
\centerline{\psfig{file=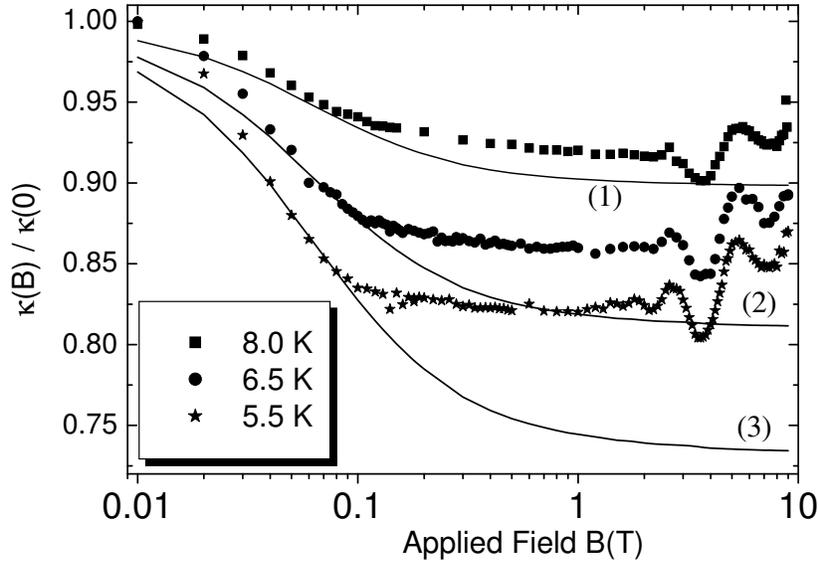,height=3.1in}}
\caption{Normalized thermal conductivity as a function of applied
field at different constant temperatures of the UC-sample measured
in this work. The continuous lines (1) to (3) are calculated using
the WF law, see Eq.(\protect\ref{wf}), and the measured in-plane
electrical resistivity at 8.0~K, 6.5~K and 5.5~K, respectively.}
\label{kappa}
\end{figure}

Let us discuss now the field dependence and its comparison with the WF law
in detail. Figure~\ref{kappa} shows the field dependence of the UC-sample
at three different constant temperatures. The observed behavior agrees
well with that published recently.\cite{ocana03} However, the measurements
for this sample show a clear kink at a field of $B \sim 0.1~$T, which is
near the field at which the MIT occurs.\cite{yakovadv} The continuous
lines (1) to (3) in Fig.~2 are obtained from
\begin{equation}
\frac{\kappa(B)}{\kappa(0)} = \frac{\kappa_e(B) - \kappa_e(0)}{\kappa(0)}
+ 1\,, \label{red}
\end{equation}
  assuming that the
phonon conductivity does not depend on magnetic field and calculating
$\kappa_e$ from Eq.(\ref{wf}). Certainly, due to the uncertainty in the
absolute values of $\kappa$ and $\rho$ the calculated curve can be shifted
up or down within experimental error. The WF curves show in
Fig.~\ref{kappa} were calculated using $\kappa(T,0)$ values that best
match the field dependence of $\kappa$ at $B < 0.1~$T; the deviation
obtained with other values of $\kappa(T,0)$ are discussed below. From
Fig.~\ref{kappa} we recognize a ``kink" in $\kappa(B)$ at a field $B \sim
0.1~$T. This represents a clear deviation from the WF law and it occurs
near the MIT, in qualitative agreement with theoretical predictions based
on the MC phenomenon.\cite{ferrer}

We note that $B \sim 0.1~$T is slightly larger than the ``critical" field
$\sim 0.07$~T obtained from scaling arguments using electrical resistivity
data.\cite{kempathesis} Measurements of the electrical resistivity as a
function of temperature at fields $0 < B < B_c$ indicate a minimum at a
temperature $T_{\rm min}(B) \propto \sqrt{B-B_c}$ (in qualitative
agreement with theory \cite{KhveshPRL2001a,GorbarPRB2002}). From these
data the ``critical" field at $T \lesssim 10~$K for the UC-sample is a
factor of two to three smaller\cite{yakovadv} than the field at which
$\kappa(B)$ shows the kink. The difference may be related to different
response of the electrical resistivity and thermal conductivity to the
induced phases and the intrinsic disorder, which may have an influence in
the formation and behavior of the ``insulating" phase. The sample disorder
potential may affect the formation of the bound states. Therefore, we
would expect that at lower temperatures, when the thermally activated
behavior of the carriers decreases, the field necessary to induce the
bound state in the electronic system increases. This is perhaps the origin
for the increase of the field at the kink measured in the AC sample below
2 K, see inset in Fig.~4(a). Note, however, that above $\sim 4~$K the
field at the kink shifts slightly to larger values the higher the
temperature, see Fig. 2, in agreement with the general behavior found
experimentally.\cite{yakovadv} Due to the relatively small QP contribution
at $T > 10~$K, the experimental resolution is not enough to resolve with
certainty the evolution of the kink position at higher temperatures.

The fact that other piece of the same UC sample shows such a clear kink
behavior might be related to the sample quality and/or smaller dimensions
compared to that one reported in Ref.~16. A less disordered sample may
enhance the sensitive of $\kappa$ to the MIT, as the results obtained for
the AC-sample indicate (see below). It may also be that the position of
the thermocouple used to measure the temperature gradient has some
influence; in this work both ends of the thermocouple pair are attached at
the same HOPG surface in contrast to the arrangement used in Ref.~16.
Experiments with different samples having different dimensions are
necessary to clarify this point.

\begin{figure}
\centerline{\psfig{file=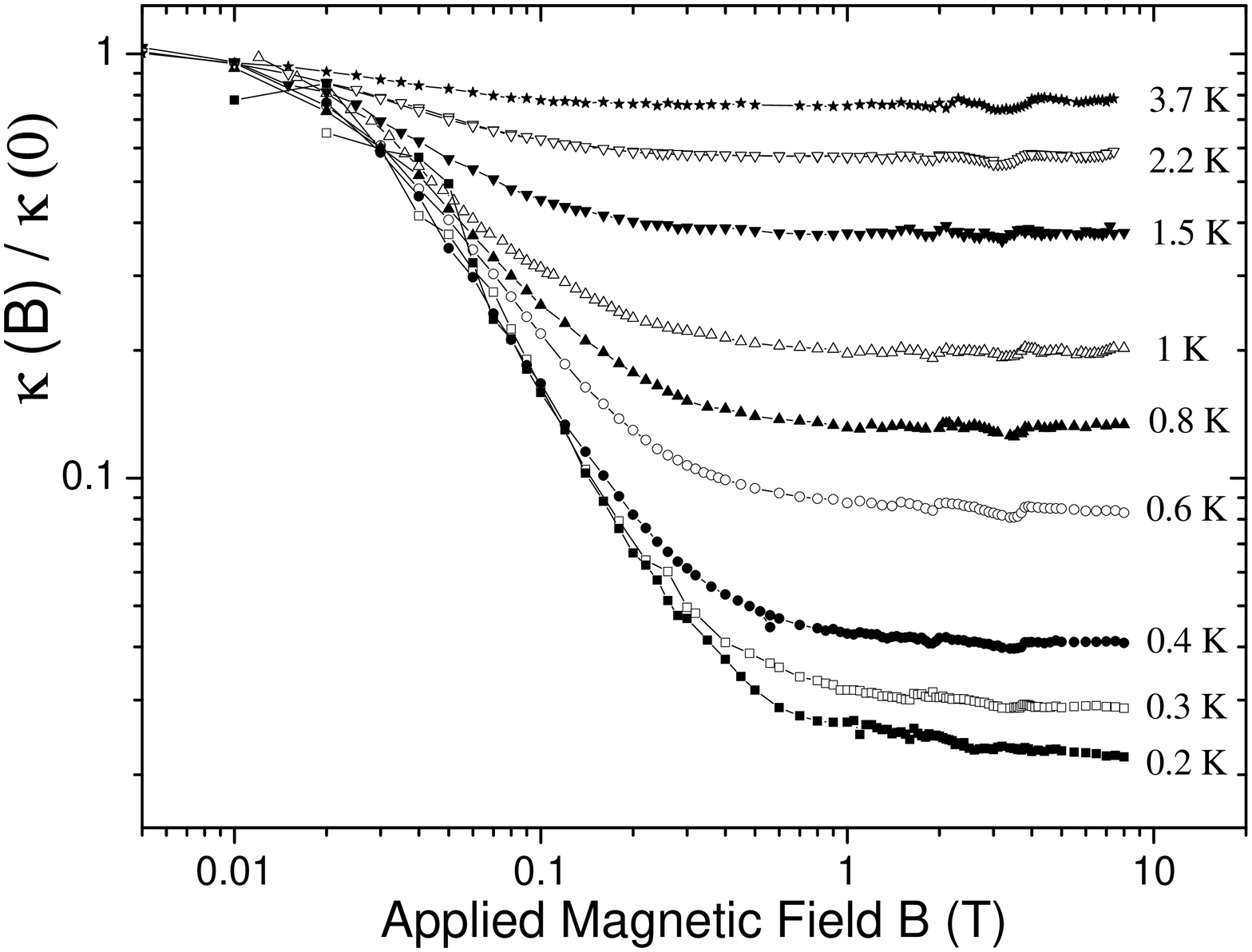,height=3.0in}} \caption{Normalized
thermal conductivity as a function of applied field at different constant
temperatures for the AC sample.} \label{klowt}
\end{figure}

\begin{figure}
\vspace{-0.8cm}\centerline{\psfig{file=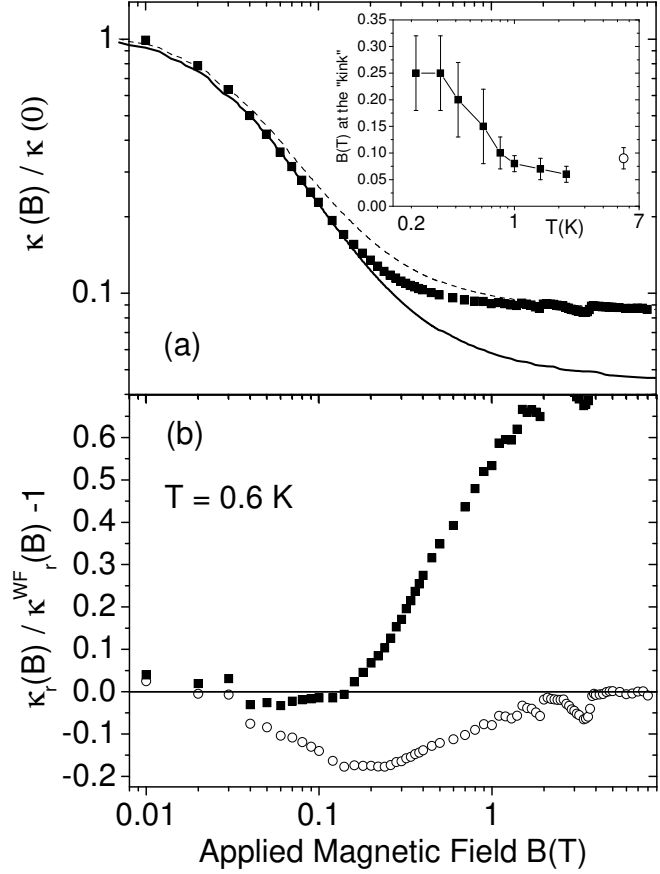,height=5.5in}}
\caption{(a) Normalized thermal conductivity as a function of
field at $T = 0.6~$K. The dashed line is obtained from
Eq.(\protect\ref{red}) choosing a value of $\kappa(0)$ that
matches the measured thermal conductivity at $B = 8~$T. The
continuous line was calculated in a similar way but $\kappa(0)$
was chosen to fit the low field range. The difference between the
two values of $\kappa(0)$ is $\sim 4\%$ and therefore within
experimental error. (b) Relative difference ($\kappa_r =
\kappa(B)/\kappa(0)$) between the measured and calculated values
from (a).
 ($\circ$) represents the values obtained from
the curve that matches the data at 8~T, and ($\blacksquare$) those
obtained from the other curve (continuous line in (a)). Inset in
(a) shows the temperature dependence of the field at which the
deviation from the WF law starts (or of the field at the minimum,
($\circ$) in (b)) for the AC sample $(\blacksquare)$. The point
$(\circ)$ is obtained for the UC sample, see Fig.~2.} \label{dif}
\end{figure}

The results for the AC sample are shown in Fig.~\ref{klowt}. This sample
is intrinsically more disordered than the UC sample as the FWHM of the
rocking curve indicates. A comparison with the WF law indicates that the
deviation starts or is at largest at the MIT. Figure ~\ref{dif}(a) shows
the same data at 0.6~K as in Fig.~\ref{klowt}. In this figure we show also
two curves calculated with Eqs.~(\ref{wf}) and (\ref{red}) using two
values of $\kappa(0.6$K,0) with a difference of $\sim 4\%$, which is
within experimental error. In one case we choose an absolute value for
$\kappa(0.6$K,0) in such a way that the WF law matches the lower field
data (continuous line in Fig.~\ref{dif}(a), as in Fig.~\ref{kappa}). In
the other case the theoretical curve based on the WF law matches the
experimental data at very high fields (dashed line in Fig.~\ref{dif}(a)).
From the difference between the experimental and theoretical curves, see
Fig.~\ref{dif}(b), we conclude that either the deviation starts at a field
near the MIT or is at largest at this field $B_c(T)$. Whatever definition
we use for the critical field, the low-temperature data below 2~K indicate
that $B_c(T)$ increases decreasing temperature, see inset in
Fig.~\ref{dif}.

\subsection{Oscillations due to Landau quantization at high
fields}

{\em (a) Quasiparticles contribution at high fields}: In what follows we
discuss  the oscillations due to Landau quantization of the electronic
levels, observed in $\kappa(B)$ at the quantum limit $(B > 1~$T) and their
possible origins. We have recently shown that these oscillations are
correlated with the QHE features observed in HOPG.\cite{ocana03} A simple
estimate indicates that the amplitude of the oscillations cannot be
explained using the WF law and the measured resistivity. The reason is
that at $B > 1~$T the resistivity increases by a factor of 100, therefore
and according to WF the electronic contribution to the thermal transport
should be much less than the phonon contribution in the temperature range
of our measurements. However, as we showed above, the WF law does neither
reproduce the field dependence of $\kappa$ at or above the MIT nor the
apparent large electronic contribution to $\kappa$ observed at high fields
and at the lowest temperatures. Therefore, we may speculate that the QP in
graphite may still have a direct contribution to the thermal transport at
the quantum limit. We note that the deviation from the WF law can be
partially explained arguing that the measured longitudinal resistivity at
large fields is not the real resistivity due to the influence of a Hall
voltage at high enough fields as has been recently suggested for
inhomogeneous semiconductors.\cite{6} The quasi-linear and non-saturating
magnetoresistance of highly oriented graphite would speak for such a Hall
contribution.

\begin{figure}
\centerline{\psfig{file=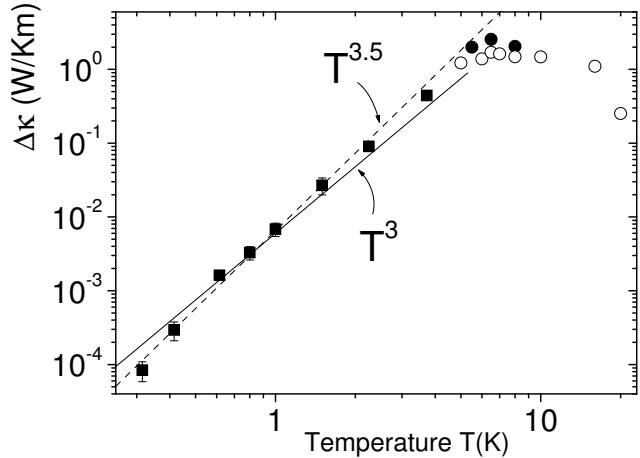,width=4.0in}} \caption{The
thermal conductivity difference between the minimum (at $B \simeq
3.7~$T) and maximum (at $B \simeq 5.5~$T) (see
Fig.~\protect\ref{kappa}) as a function of temperature for
$(\blacksquare$) the AC sample, ($\bullet$) the UC sample studied
in this work and $(\circ$) the UC sample from Ref.~16. The
continuous and dashed lines show a $T^3$ and $T^{3.5}$ temperature
dependence, respectively.} \label{deltak}
\end{figure}

As a characterization of the oscillations we have defined in Ref.~16 the
oscillation amplitude $\Delta k = \kappa(B \simeq 3.7~$T)$- \kappa(B
\simeq 5.5~$T) and showed that $\Delta \kappa$ has an apparent maximum
around 7~K. Figure~\ref{deltak} shows those data with new points obtained
for the other piece of the same sample. The oscillation amplitude was
measured in the AC sample down to 0.2~K. The data indicate that the
oscillation amplitude vanishes at low temperatures following roughly a
$T^3$ or a $T^{3.5}$ law below $\sim 5~$K, see Fig.~\ref{deltak}. Because
the QHE features are clearly measured below 1~K,\cite{yakovprl}  this
$T$-dependence suggests that the oscillations might be related to
phonon-mediated transitions, i.e. either the oscillations are due to a
change in the phonon mean free path through their interaction with the QP
or the QP contribution to the transport, being not as negligible as the WF
law predicts, is influenced by the scattering of QP with phonons. Another
interesting behavior that is not expected from the WF law is the steadily
increase (leaving the oscillations of $\kappa$ by side) of the mean value
of $\kappa(B> 1$T$, T \sim 5$K$)$. This increase with field is
reproducible in all measured samples and is at largest at the temperature
where the oscillation amplitude is at maximum, see Fig.~5. Taking into
account the experimental and theoretical research of the anomalous
properties of graphite, the still unclear effects of the linear dispersion
relation for the QP including Landau quantization and phonon-QP
interaction in the theory for thermal transport and the unclear origin of
the linear magnetoresistance, at the present stage we are not able to
provide a conclusive answer on the QP contribution to the thermal
transport at the quantum limit.

{\em (b) Phonon-electron interaction}: Due to the correlation of the
oscillations with the Hall effect, it is clear that the QP play a role. As
in Antimony,\cite{long} we may argue that phonon-electron interaction is
the main scattering mechanism for phonons that leads to the oscillations
in the phonon contribution of $\kappa$, through an effect of the
quantization on the phonon-electron scattering, due to, e.g. the field
dependence of the density of states (or carrier density) of the QP.
However, the relatively low density of QP ($10^{-4}$ to $10^{-5}$ carriers
per atom) indicates that this scattering mechanism in graphite should be
less important than in usual metals. The inelasticity parameter $\eta = v
/\lambda \omega_c$ (here $v$ is the sound velocity, $\lambda$ the magnetic
length and $\omega_c$ the cyclotron frequency), that provides an estimate
of the efficiency of the phonon-electron scattering, is $\sim 0.01$ at 4~T
for graphite, apparently too small to indicate a strong phonon-electron
interaction. From the other electronic side, the thermally activated
$T-$dependence of the longitudinal electrical resistivity obtained for
graphite at low magnetic fields, also observed in several 2D electron
systems \cite{yakovnar}, and the $T-$dependence for $\kappa$ do not speak
for a large electron-phonon interaction.

It may be possible that at certain magnetic fields the phonon-electron
scattering is significantly enhanced when the separation energy between
Landau levels is of the order of the characteristic acoustic-phonon
energy.\cite{golo} This is expected to occur at a temperature $T_\lambda
\sim \hbar \omega = \hbar v q$ when the phonon wave vector is of the order
$1/\lambda$, the inverse of the magnetic length. Putting numbers one
obtains that $T_\lambda \sim 2~$K at $B = 3.5~$T, surprisingly near the
maximum temperature of the oscillation amplitude. However, energy
conservation requires that the phonon energy matches the separation
between Landau levels. At $B = 5~$T, $\hbar \omega_c \sim 100~$K taking
the effective mass of electrons $m^* \sim 0.05 m_0$,\cite{soule} and
therefore much larger than $T_\lambda$, making the one-phonon electron
transition inefficient. Multi-phonon processes may, however, provide the
necessary efficiency for that transition to occur.\cite{golo} One may also
argue that the low effective mass of the carriers used above to estimate
$\hbar \omega_c$ is not the appropriate one, since in this field range
electron-hole excitonic pairs may act as the main carriers.

Other possibility for a resonance phonon-electron scattering is a
substantial decrease of the energy between Landau levels due to the
intersection of two Landau levels belonging to different size-quantization
subbands (due to electrons and holes, for example). There is however no
clear experimental evidence for such a crossover. Some details of the
field dependence of the Hall effect depend on sample quality and its
internal disorder, and a rather complicated algorithm is used to obtain
details of the band structure from these data. Very recently, clear
quantized plateaus were obtained for the Hall conductivity in two
different geometries in high quality and small HOPG
samples.\cite{kempathesis} If this behavior represents that of ideal
graphite, then some characteristics of the carriers and the electronic
band structure picture of graphite reported in the past should be revised.

When discussing the efficiency of the phonon-electron scattering in
graphite we should take into account
 an effective Debye
temperature $\theta^*$. This effective temperature provides the limiting
value for the change of the scattered phonon wave vector $q$. Because of
the low density of carriers $2 k_F \ll q_D$ the effective Debye
temperature for the phonon-electron interaction is estimated as $\theta^*
\sim 2 k_F v \hbar /k_B \sim 9~$K  using the 2D value for $2 k_F$ obtained
from Hall effect measurements on the same sample.\cite{kempathesis} This
value should be compared with the (anisotropic) Debye temperature for
graphite with a value $\theta > 500~$K.\cite{kel} This effective Debye
temperature $\theta^*$ is near the temperature at which we observe a
maximum in the oscillation amplitude, see Fig.~\ref{deltak}.

\section{Summary}

In summary, we have shown that the magnetic field dependence of the
thermal conductivity in highly ordered graphite shows a kink anomaly at or
near the ``critical" field of the metal-insulator transition. Above this
field a clear plateau is measured. The observation of this kink depends on
sample; experimental evidence suggests that the larger the disorder the
less clear is the kink, but more systematic work is necessary. For the two
measured samples we have shown that the WF law does not account for the
field dependence of $\kappa$ showing a systematic deviation at the
metal-insulator transition. The overall behavior appears to be accounted
for by the magnetic catalysis model, the opening of an excitonic gap and
the increase of the effective mass of the carriers due to the
electron-hole pairing. We have shown also that the WF law does not account
for the oscillations in $\kappa$ due to Landau quantization at the quantum
limit. Measurements to 0.2~K show that the oscillation amplitude increases
with temperature below $\sim 5~$K with a power law $T^n$ with $n = 3.0
\ldots 3.5$ suggesting that phonons play a role.

\section*{ACKNOWLEDGMENTS}
Fruitful discussions with  V. de la Incera, E. Ferrer, Y. Kopelevich, M.
A. Vozmediano and S. G. Sharapov are gratefully acknowledged. We thank M.
Ramos for his kind assistance in the calibration of the RuO$_2$
thermometers.


\begin{thebibliography}{9}

\bibitem{neckel86}H. Neckel, P. Esquinazi, G.  Weiss  and  S.  Hunklinger,
Solid State Commun. {\bf 57}, 151 (1986).

\bibitem{esqui04}P. Esquinazi, M. Ramos and R. K\"onig, J. Low Temp. Phys.
{\bf 135}, 27 (2004).

\bibitem{kel}B. T. Kelly,  {\it Physics of Graphite}, p. 267 ff,
p. 293, p. 322 ff. Applied Science Publishers LTD, London and New Jersey
(1981).

\bibitem{dre}M. S. Dresselhaus, G. Dresselhaus, K. Sugihara, I. Spain,
and H. A. Goldberg, in
 {\it Graphite Fibers and Filaments},
Springer Series in Material Science Vol. 5, pp. 179-188 (Springer-Verlag,
1989).

\bibitem{semenoff}G. W. Semenoff, Phys. Rev. Lett. {\bf 53}, 2449 (1984).

\bibitem{divin}D. P. Di Vicenzo and E. J. Mele, Phys. Rev. B {\bf 84},
1685 (1984).

\bibitem{gonzalez}J. Gonz\'alez, F. Guinea and M. A. Vozmediano, Nucl.
Phys. B {\bf 406}, 771 (1993); Phys. Rev. B {\bf 63}, 134421 (2001).

\bibitem{tsuei}C. C. Tsuei and J. R. Kirtley, Rev. Mod. Phys. {\bf 72},
969 (2000).

\bibitem{aya}C. Ayache, Ph.D. Thesis, Grenoble 1978 (unpublished). One of
the curves of the thesis can be found in {\it Landolt-B\"ornstein}, New
Series III/15c, page 433 (1991).

\bibitem{woo}J. A. Woollam, Phys. Rev. B {\bf 3}, 1148 (1971).

\bibitem{yakovprl} Y. Kopelevich,  J. H. S. Torres, R. R. da Silva,
F. Mrowka, H. Kempa and P.
Esquinazi, Phys. Rev. Lett. {\bf 90}, 156402 (2003).

\bibitem{kempassc}H. Kempa, Y. Kopelevich, F. Mrowka,
A. Setzer, J. H. S. Torres, R. H\"ohne, and P. Esquinazi, Solid State
Commun. {\bf 115}, 539 (2000).

\bibitem{ferrer} E.J. Ferrer, V.P. Gusynin, V. de la Incera, Modern Physics
Letters B {\bf 16}, 107 (2002); Eur. Phys. J. B {\bf 33}, 397 (2003).

\bibitem{KhveshPRL2001a} D. V. Khveshchenko, Phys. Rev. Lett. {\bf87}, 206401 (2001);
ibid. {\bf87}, 246802 (2001).

\bibitem{GorbarPRB2002} E. V. Gorbar, V. P. Gusynin, V. A. Miransky, and
I. A. Shovkovy, Phys. Rev. B {\bf66}, 045108 (2002).

\bibitem{ocana03}R. Oca\~na, P. Esquinazi, H. Kempa, J. Torres, and Y.
Kopelevich, Phys. Rev. B {\bf 68}, 165408 (2003).

\bibitem{KopelPSS1999} Y. Kopelevich, V. V. Lemanov, S. Moehlecke, and J. H. S. Torres,
Phys. Solid State {\bf41}, 1959 (1999); Fiz. Tverd. Tela (St.
Petersburg) {\bf 41}, 2135 (1999).

\bibitem{SercheliSSC2002} M. S. Sercheli, Y. Kopelevich,
R. Ricardo da Silva, J.H.S. Torres, and C. Rettori, Solid State Commun.
{\bf 121}, 579 (2002).

\bibitem{KempaPRB2002} H. Kempa, P. Esquinazi and Y. Kopelevich,
Phys. Rev. B {\bf 65}, 241101(R) (2002).

\bibitem{AbrahamsRMP2001} E. Abrahams, S. V. Kravchenko, and M. P. Sarachik,
Rev. Mod. Phys. {\bf 73}, 251 (2001), and references therein.

\bibitem{GusyninPRL1994} V. P. Gusynin, V. A. Miransky, and I. A. Shovkovy,
Phys. Rev. Lett. {\bf 73}, 3499 (1994).

\bibitem{yakovadv}Y. Kopelevich,  P. Esquinazi, J. H. S. Torres, R. R. da
Silva, and H. Kempa, Advances in Solid State Physics {\bf 43}, 207
(2003).


\bibitem{krishana2}K. Krishana, N. Ong, Q. Li, G. Gu, and N.
Koshizuka, Science {\bf 277}, 83 (1997).

\bibitem{aubin}H. Aubin et al., Phys. Rev. Lett. {\bf 82}, 624 (1999).

\bibitem{pogo} Yu. Pogorelov, M. Arranz, R. Villar, and S. Vieira,
Phys. Rev. B {\bf 51}, 15474 (1995).

\bibitem{talden}A. N. Taldenkov, P. Esquinazi and K. Leicht, J. Low Temp.
Phys. {\bf 115}, 15 (1999).

\bibitem{ando} Y. Ando, J. Takeya, Yasushi Abe, K. Nakamura and A.
Kapitulnik, Phys. Rev. B {\bf 62}, 626 (2000).

\bibitem{vief}V. de la Incera and E. Ferrer, cond-mat/0309046 and refs.
therein.

\bibitem{yang}X. Yang and C. Nayak, Phys. Rev. B {\bf 65}, 064523 (2002).



\bibitem{iny} A. Inyushkin,  K. Leicht, and P. Esquinazi, Cryogenics {\bf 38}, 299 (1998).

\bibitem{ulrich}K. Ulrich, Diplomarbeit, University of Leipzig,
2004, unpublished.


\bibitem{oca2}R. {Oca\~{n}a} and  P. Esquinazi, Phys. Rev. B {\bf 66},
064525 (2002).

\bibitem{ruo2}B. Neppert
and P. Esquinazi, Cryogenics  {\bf 36}, 231 (1996).

\bibitem{morelli}D. T. Morelli and C. Uher, Phys. Rev. B {\bf 31}, 6721
(1985).

\bibitem{am}See, for example, N. W. Ashcroft and N. D. Mermin, in {\it Solid State Physics},
Holt-Saunders International Editions, 1976.


\bibitem{car}M. Inagaki, in {\it New carbons}, Elsevier Science Ltd. 2000, page 52.


\bibitem{chau}C. K. Chau and S. Y. Liu, J. Low Temp. Phys. {\bf 15} 447
(1974).

\bibitem{kempathesis}H. Kempa, Ph. Thesis, University of Leipzig (2004),
unpublished.

\bibitem{6}M. M.  Parish and P. B. Littlewood, Nature {\bf 426}, 162 (2003).

\bibitem{long}J. R. Long, C. G. Grenier, and J. M. Reynolds, Phys.
Rev. B {\bf 140}, A187 (1965).

\bibitem{yakovnar}Y. Kopelevich, P. Esquinazi,J. H. S. Torres, R. R. da
Silva, H. Kempa, F. Mrowka and R. Oca\~na, in {\it Studies of High
Temperature Superconductors}, A. Narlikar (ed.), NOVA Science Publishers,
Inc. (2003),  Vol. 45, page 71.

\bibitem{golo}V. N. Golovach and M. E. Portnoi, cond-mat/0202179.

\bibitem{soule}D. E. Soule, Phys. Rev. {\bf 112}, 708 (1958).

\end{thebibliography}
\end{document}